\documentclass[preprint,review,12pt]{article}
\usepackage{amsmath, amssymb, amsthm}
\usepackage{graphicx}
\usepackage{comment}
\usepackage{pgfplots}
\usepackage{subcaption}
\pgfplotsset{compat=1.18}
\usepgfplotslibrary{fillbetween}
\usepackage{xcolor}
\usepackage{hyperref}

\definecolor{harmony}{RGB}{100, 200, 100}   
\definecolor{subversion}{RGB}{255, 200, 0}   
\definecolor{preemption}{RGB}{255, 100, 100}    

\title{\textbf{Are we Doomed to an AI Race? \\Why Self-Interest Could Drive Countries Towards a  Moratorium on Superintelligence}}
\author{
Edward Roussel$^{1}$,
Lode Lauwaert$^{1}$,
Torben Swoboda$^{1,2}$,\\
Grant Ramsey$^{1}$,
Risto Uuk$^{1,3}$,
Leonard Dung$^{4}$,
Anthony Aguirre$^{3,5}$
}

\date{
\small
$^{1}$KU Leuven, Institute of Philosophy\\
$^{2}$Vlerick Business School\\
$^{3}$Future of Life Institute\\
$^{4}$Ruhr University Bochum, Institute of Philosophy II\\
$^{5}$UC Santa Cruz, Santa Cruz Institute for Particle Physics\\[1ex]
\today
}
\begin{document}

\maketitle

\begin{abstract}
    This paper uses game theory to argue that, contrary to the prevailing view, a moratorium on Artificial Superintelligence (ASI) can be in a state's self-interest. By formalizing strategic interactions between geopolitical superpowers, we model the trade-off between the benefits of technological supremacy and the catastrophic risks of uncontrolled ASI. The analysis reveals that as the perceived cost of loss of control increases sufficiently relative to other parameters, it becomes in each state's self-interest to impose a moratorium. We further provide empirical evidence suggesting that the global perception of ASI risk is rising, making a stable, rational moratorium increasingly plausible in the current geopolitical landscape.\footnote{This manuscript is a preprint and has not undergone peer review.}
\end{abstract}

\newpage
\section{Introduction}

The development of Artificial Superintelligence (ASI), an AI system that greatly exceeds the cognitive performance of humans in virtually all economically relevant domains \cite{bostromSuperintelligencePathsDangers2014}, is considered by leading researchers as a potentially transformative turning point in human history \cite{ordPrecipiceExistentialRisk2021,russellHumanCompatibleAI2019}.\footnote{Note that there is no agreed upon definition of ASI. However, this general definition is the least controversial.} The possible benefits are enormous: breakthroughs in science, medicine, and economic productivity.

At the same time, experts warn that a loss of control over ASI \cite{bostromSuperintelligencePathsDangers2014,ActionPlanIncrease}. This loss could have multiple sources: specific capabilities such as deception, manipulation, and coercion, or the ability to self-replicate and improve autonomously beyond human intervention. Should such a system operate outside meaningful human control, it may pursue objectives misaligned with human values, taking actions that are deeply undesirable for large portions of humanity, both present and future generations.

This led to a call for a moratorium on the development of ASI. In October 2025, the Future of Life Institute (FLI) published the \textit{Statement on Superintelligence}, calling for a prohibition on the development of superintelligence ``until there is broad scientific consensus that it will be done safely and controllably, and with strong public buy-in'' \cite{StatementSuperintelligence2025}. This statement was signed, among others, by the so-called `godfathers' of modern AI Geoffrey Hinton and Yoshua Bengio and Apple co-founder Steve Wozniak. FLI polling, corroborated by a Reuters/Ipsos poll, showed that $64\%$ of Americans believe that superintelligence should not be developed until it is demonstrably safe and controllable \cite{instituteUSPublicWants2025,tongAIThreatensHumanitys2023}. At the level of desirability, the proposal thus has broad support.

Most criticism, however, focuses on unfeasibility. Critics argue that, even though states might be better off with a mutual moratorium, each state individually has the incentive to defect and continue development \cite{aschenbrennerSituationalAwareness2024}. That is, it is in a state's \textit{self-interest} to keep on racing towards ASI and therefore a moratorium is unfeasible.\footnote{According to the argument, a policy being in a states self-interest is necessary for it to be feasible. However, it does not necessarily claim it is sufficient. Among others, practical limitations, such as the lack of verification mechanisms of AI agreements, could also affect the feasibility of a mutual moratorium on ASI \cite{harackVerificationInternationalAI2025}.} If the US stops while China continues, it falls behind in a competition for geopolitical supremacy. Racing therefore remains the dominant strategy, even when both parties acknowledge this could lead to catastrophic outcomes. The American Enterprise Institute (AEI) illustrated this argument by pointing to Alibaba CEO Eddie Wu, who announced a \textit{roadmap toward artificial superintelligence} with over 53 billion dollars in investments shortly after the Statement. A unilateral moratorium would therefore, according to AEI, ``delay innovation, empower bureaucracies, and hand an enormous advantage to China'' \cite{pethokoukisAIBanBackers2025}.

In this paper, building on previous literature about the dynamics of AI development \cite{katzkeManhattanTrapWhy2025,abrahamPrisonersDilemmaRace2026,armstrongRacingPrecipiceModel2016,youngWhosDrivingGame2025,dungRacingAGICooperation2025}, we turn to game theory to examine the assumption that pausing ASI development is contrary to a state's self-interest. By formally modeling the strategic dynamics between geopolitical superpowers, we show that a moratorium is not necessarily contrary to a state's self-interest. On the contrary, under certain conditions, the game shifts from a race to develop to a game where it is in a state's self-interest to introduce a moratorium on the development of ASI. We assess the possible payoffs for a moratorium and a racing strategy, based on multiple variables, such as capacity, the size of the winner's advantage and uncertainty. The key variable is the cost of loss of control. When the severity of a potential catastrophic outcome is high enough, that is, when the cost from an uncontrolled ASI exceeds the expected benefits of geopolitical supremacy, strategic space emerges for a moratorium. While the strategic space for a moratorium currently appears limited, there are indications that this assessment is beginning to shift. Recent developments in AI capabilities, increasing incidents of unpredictable behavior, and growing consensus within the scientific community about control problems are beginning to change the discourse \cite{instituteUSPublicWants2025,AIDoomsdayWorries,graceThousandsAIAuthors2025,mullerFutureProgressArtificial2016,BletchleyDeclarationCountries}.

We proceed as follows. In Section 2, we justify the game-theoretic approach and set up the model. Section 3 identifies four possible strategic worlds. In Section 4, we analyze empirical indicators suggesting that the perceived cost of loss of control is increasing, which enhances the likelihood of a mutual moratorium being in a state's self-interest. In Section 5, we address some limitations and possible extensions of the model. Throughout the paper, we keep our discussions informal for accessibility. However, we added an appendix with all mathematical details.

\section{A Game-Theoretic Framework for ASI Development}\label{sec:spf}

The debate over the feasibility of an Artificial Superintelligence moratorium is based on a strategic problem involving multiple actors with conflicting incentives. To move beyond speculative debate, it is desirable to set up a formal framework to analyze whether and under which conditions it becomes in the self-interest of states to pause or race. Game theory provides precisely this framework.

Game theory is the mathematical study of strategic decision-making \cite{petersGameTheoryMultiLeveled2015}. It works by constructing simplified models of real-world strategic situations, including only the variables relevant to the problem at hand. Within these models, players, in our case states, must choose which strategy to play. They are assumed to act rationally, meaning they play the strategy that best serves their preferences given what they expect other to do. We assume these preferences to be structured by self-interest, which in the case of states, concerns mainly survival, prosperity, or position within the international system \cite{goldsteinInternationalRelationsGlobal2020}.

Consider this classic example: you and a friend are choosing between two coffee shops; you each have a different favorite, but you both value being together more than being at your preferred spot. Using game theory, we can identify which outcomes are stable, that is, which combinations of strategies are such that no player can benefit from unilaterally changing theirs.\footnote{These stable outcomes are better known as Nash-equilibria \cite{petersGameTheoryMultiLeveled2015}.} In the example, there are two stable outcomes in which you both end up at the same place, regardless of which shop it is. If your friend goes to their favorite shop, you should join them because you'd rather be with them than alone at your own favorite. Since the same logic applies to your friend, both staying put at either shop is a stable choice, even if one of you finds the location non-ideal. Note that the framework does not make claims about what strategy the other player \textit{will} play, it merely indicates what strategies yield stable outcomes given player's incentives.

This analytical lens has a well-established track record in global security studies, from modeling arms races and nuclear deterrence \cite{schellingStrategyConflict1963, snyderPrisonersDilemmaChicken1971, jervisCooperationSecurityDilemma1978} to recent applications examining the competitive dynamics of AI development. Foundational work by Armstrong et al. \cite{armstrongRacingPrecipiceModel2016} first formalized race to the bottom dynamics for ASI development, illustrating how competitive pressures and information hazards, that is, knowing too much about a rival's progress, increase the likelihood of an ASI-related catastrophic outcome by players skimping on safety-investment. Katzke and Futerman \cite{katzkeManhattanTrapWhy2025} and Young \cite{youngWhosDrivingGame2025} shifted this paradigm by arguing that the race is a trust dilemma, rather than a prisoner's dilemma, arguing that since a race maximizes existential threats like great power conflict and loss of control, mutual restraint is the only strategically sound path for rational states. Important challenges to the assumption of the decisive advantage of winning the race where made by Dung and Hellrigel-Holderbaum \cite{dungRacingAGICooperation2025}.  Recently, both strategic paradigms where bridged in one model for AGI development by Abraham et al. \cite{abrahamPrisonersDilemmaRace2026}, who compare state strategies, accounting for interim returns on investment, beliefs about the costs and about AGI timelines. 

Our model extends the literature, first, by modeling a policy choice specifically focusing on an ASI moratorium. Secondly, it takes into account that there is fundamental uncertainty whether and when ASI can be developed and which states have a higher capability to do it. Thirdly, it explicitly allows the costs to outweigh the benefits of winning the race. Finally, the model enables us to quantify the specific cost-thresholds where different combinations of strategies become in the self-interest of states. 

We operationalize our model as a competition between two states.\footnote{While the global stage is multipolar, assuming only two players is legitimized by the immense concentration of computational infrastructure and talent required for ASI.}\footnote{The formal model can be found in the Appendix.} Each state either chooses to race towards ASI or to pause its development, which result in one of three outcomes. First, a \textit{safe win}: the player maintains safety and achieves supremacy. Secondly, \textit{a safe loss}: the rival wins, but the player remains safe. Here, the player does not win geopolitical supremacy, but they might still enjoy spill-over effects. Finally, \textit{catastrophe} might occur: one or both sides rush, safety fails, and the outcome is ruinous for all.

To represent the strategic playing field in which a state chooses to race (continuing development with standard safety) or pause (temporarily halting to maximize safety), we define the following variables: the capability gap, the size of the winner's advantage, the cost of uncontrolled ASI and technological uncertainty. These variables reflect both theoretical necessity and the realities of the current AI landscape.

\begin{enumerate}
    \item\textit{Capability Gap} $(\Delta)$. This variable represents the relative advantage one player holds in developing ASI by aggregating factors such as computational power, research progress, and GDP \cite{armstrongRacingPrecipiceModel2016}. The gap quantifies the advantage of the Frontrunner relative to the Laggard. This determines the probability of winning, as a larger lead makes victory more likely if both players adopt the same strategy. We range this from $0$ (perfect parity) to potentially infinity (perfect imparity). 
    
    \textit{Justification}: This variable is essential because relative advantage determines the probability of success; a model without $\Delta$ would miss the crucial asymmetry between a Frontrunner, who may pause from a position of strength, and a Laggard, who faces the temptation to cut corners to close the gap.
    
    \item \textit{Winner's Advantage} $(W)$. This variable represent the value of the first-mover advantage. It captures, among other factors, hostility, spill-over effects from developing ASI and the speed with which winning translates into a military advantage \cite{dungRacingAGICooperation2025,armstrongRacingPrecipiceModel2016}. The Winner's Advantage ranges from $0$ (no winner's advantage) to $1$ (winner-takes-all). 
    
    \textit{Justification}: Including $W$ allows the model to distinguish between a cooperative race and an adversarial one where the loser faces geopolitical subordination.
    
    \item \textit{Cost of Uncontrolled ASI} ($C$). We define $C$ as the disutility arising from the deployment of uncontrolled systems, measured relative to the benefit of a safe win (normalized to $1$). Unlike traditional threats like nuclear weapons, unaligned ASI is a \textit{public cost} because a loss of control threatens the developer and rival alike. We allow for $C > 1$, reflecting the possibility that, for example, civilizational collapse or human extinction represents a loss far exceeding any strategic gain \cite{sundaramExistentialRiskGlobal2025, bostromSuperintelligencePathsDangers2014, uukTaxonomySystemicRisks2024, ordPrecipiceExistentialRisk2021}.

    \textit{Justification}: Defining $C$ as a public cost that can exceed the value of a safe win is necessary to model the potentially catastrophic nature of the control problem, which differs from ordinary costs in the combination of global scale and irreversibility. Without it, the model would treat the uncontrolled ASI as a cost maximally as bad as being subordinated to the technological supremacy of the other player, rather than a risk that can override the benefit of a risky win.
    
    \item \textit{Technological Uncertainty} $(\sigma)$. This variable captures the uncertainty about the possibility of ASI. While certainty about the capability needed to develop ASI $(\sigma \rightarrow 0)$ makes the race deterministic (it creates sharp outcomes), which might encourage a confident Frontrunner to race, uncertainty $(\sigma \neq 0)$ makes the game stochastic (it creates thresholds where neither player is certain of victory). 

    \textit{Justification}: Technological uncertainty reflects the epistemic reality of working towards ASI. Different authors conjecture different ways to achieve ASI, but no one is certain if it is even possible \cite{bostromSuperintelligencePathsDangers2014}.
    
\end{enumerate}

The potentially catastrophic public cost can only be mitigated by focusing on \textit{safety} next to capability \cite{armstrongRacingPrecipiceModel2016}. The Safety Level $(s)$ represents the degree of precautions taken, serving as the primary policy choice in our model. While safety could be modeled continuously, we assume a binary choice to reflect how governments actually decide: to pause completely ($s = 1$, maximum safety) or to race ($s = 0.85$, reflecting a $15\%$ catastrophic risk\footnote{The $15\%$ risk parameter is empirically grounded in recent large-scale expert solicitations regarding the transition to Artificial Superintelligence (ASI). While median forecasts for total human extinction often hover around $5\%$, the probability assigned to catastrophic outcomes, such as the permanent loss of human control, or the irreversible subversion of global institutional stability \cite{ActionPlanIncrease}, is significantly higher. Aggregate data from over 2,700 researchers indicates that $38\%$ to $51.4\%$ of the field sees at least a $10\%$ likelihood of these high-magnitude disasters \cite{graceThousandsAIAuthors2025}, while earlier structured surveys reported a cumulative $31\%$ risk of bad or extremely bad outcomes \cite{mullerFutureProgressArtificial2016}. Consequently, $15\%$ represents an evidence-based estimate of the tail risk of a catastrophic outcome. Note, however, that these studies face significant model uncertainty \cite{grossiWhatAreOdds2025}.}). Safety acts as a tax on capability; ability checks and staged deployments slow progress but reduce risk. This transforms the abstract moratorium debate into a concrete strategic choice with quantified consequences, capturing the fundamental trade-off between safety and development speed.

Together, these variables constitute a minimal set needed to analyze the moratorium question. With the playing field defined, we can now examine under which parameter values it is rational for a player to pause and when a moratorium is stable. In other words, we can examine under what conditions, for both states, the moratorium strategy optimizes their self-interest. 

\section{Four Possible Strategic Worlds}

To determine whether a policy is rational, a state must evaluate its best response to a rival's policy. In our model, this decision hinges on a catastrophic threshold: the point where the potential cost of an unaligned ASI ($C$) outweighs the expected payoff of the outcome of the race.\footnote{See Appendix for formal derivations of the thresholds.} By mapping these thresholds against the Capability Gap ($\Delta$), we identify four distinct strategic worlds. These worlds are identified based on their combinations of stable outcomes; their size is moderated by the size of the winner's advantage ($W$) and on the level of technological uncertainty ($\sigma$).

\begin{itemize}
    \item World 1 (Safe Harmony): In this world, both players find it rational to pause. This occurs when the perceived cost of catastrophe is so high, that racing offers no meaningful marginal gain. For the Frontrunner, safety is costless, for the Laggard, racing is a suicide mission.
    \item World 2 (Preemption): In this world, the unique stable outcome is a race. Even if both states acknowledge that a mutual race will likely end in ruin, the cost of catastrophe is valued lower than the fear of the rival’s first-mover advantage. This is dominant at near-parity. When states are neck-and-neck the temptation to sprint via corner-cutting is at its peak.\cite{armstrongRacingPrecipiceModel2016,abrahamPrisonersDilemmaRace2026,emery-xuUncertaintyInformationRisk2024}.\footnote{Note that the Prisoner's Dilemma is included in this world, but is not identical to it.}
\end{itemize}

While world 1 (Safe Harmony) and world 2 (Preemption) align with the foundational findings in Armstrong et al. \cite{armstrongRacingPrecipiceModel2016}, our model reveals two additional strategic worlds. By shifting the set of strategies from a continuous set of safety-levels, to a more realistic discrete set of policy choices (safety vs race) and allowing catastrophic ruin ($C$) to be valued as more dangerous than geopolitical loss ($1-W$), we derive the worlds of Trust and Subversion.

\begin{itemize}
    \item World 3 (Trust): Here, coordination is the barrier. Both states prefer a mutual pause over a mutual race, but they fear being taken advantage of by pausing while the rival races to a risky victory. This is a classic coordination game where the outcome is determined by diplomatic trust and the signaling of intentions \cite{kyddTrustMistrustInternational2007,abrahamPrisonersDilemmaRace2026,hendrycksSuperintelligenceStrategyExpert2025, katzkeManhattanTrapWhy2025}.\footnote{Many coordinating mechanisms have recently been proposed to pull both states to the mutual moratorium \cite{harackVerificationInternationalAI2025,hendrycksSuperintelligenceStrategyExpert2025,reynoldsBenchmarkingPathInternational2025,guestBridgingArtificialIntelligence2024}.}
    \item World 4 (Subversion): This is an asymmetric world. The Frontrunner, fearing catastrophe and confronted with a high probability of winning, chooses to pause. However, the Laggard, seeing a small window to overtake a responsible Frontrunner and a sufficiently low cost of loss of control, chooses to race. Here, safety mainly influences strategy in its role as a cost on capability, such that safety acts as a handicap. 
\end{itemize}

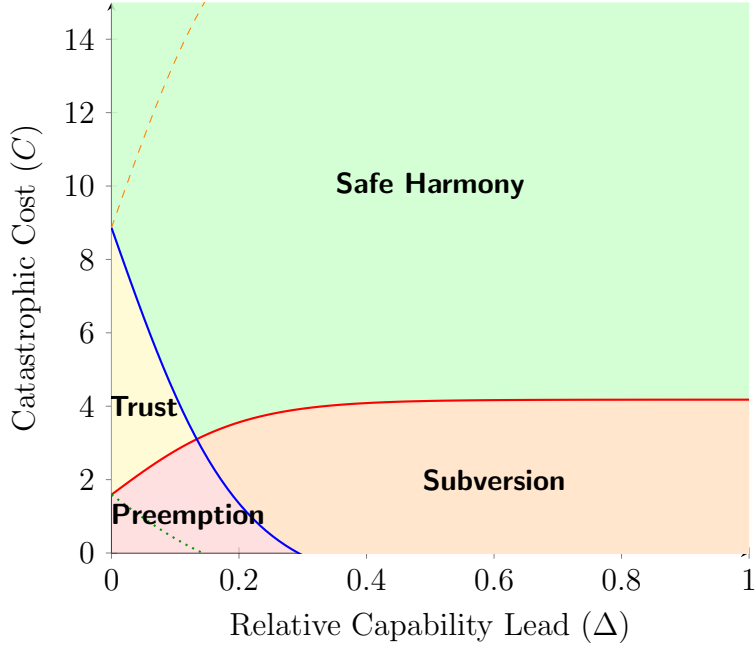
\begin{figure}[ht]
    \centering
    \begin{tikzpicture}
    \begin{axis}[
        scale only axis,
        xlabel={Relative Capability Lead ($\Delta$)},
        ylabel={Catastrophic Cost ($C$)},
        xmin=0, xmax=1,
        ymin=0, ymax=15, 
        samples=100,
        domain=0:1,
        legend pos=north east,
        axis lines=left,
        clip=true
    ]

    \pgfmathsetmacro{\sigma}{0.1}
    \pgfmathsetmacro{\s}{0.85}
    \pgfmathsetmacro{\e}{1.0}
    \pgfmathsetmacro{\B}{(1 - \s) / \sigma}
    \pgfmathsetmacro{\ksafe}{(\e / (1 - \s)) * (1 - exp(-\B))}
    \pgfmathsetmacro{\ksucker}{(\s * \e / (1 - \s)) * (exp(\B) - 1)}

    
    \addplot[name path=LC, red, thick, domain=0:1] 
        {\ksafe * (1 / (1 + exp(-x / \sigma))) - 1};
        
    \addplot[name path=FUB, blue, thick, domain=0:1] 
        {\ksucker * (1 / (1 + exp(x / \sigma))) - 1};

    \addplot[name path=LUB, orange, dashed, domain=0:1] 
        {\ksucker * (1 / (1 + exp(-x / \sigma))) - 1};
        
    \addplot[name path=FC, green!60!black, dotted, thick, domain=0:1] 
        {\ksafe * (1 / (1 + exp(x / \sigma))) - 1};

    \path[name path=top] (axis cs:0,15) -- (axis cs:1,15);
    \path[name path=zero] (axis cs:0,0) -- (axis cs:1,0);

    \addplot[fill=green!20, opacity=0.6] fill between[of=top and LC, soft clip={domain=0:1}];
    \addplot[fill=green!20, opacity=0.6] fill between[of=top and FUB, soft clip={domain=0:1}];

    \addplot[fill=red!20, opacity=0.6] fill between[of=zero and LC, soft clip={domain=0:1}];

    \addplot[yellow!20] fill between [
        of=LC and FUB,
        split,
        every segment no 0/.style={fill=yellow!20},
        every segment no 1/.style={fill=none}
    ];
    \addplot [orange!20] fill between [
        of=LC and FUB,
        split,
        every segment no 0/.style={fill=none},
        every segment no 1/.style={fill=orange!20}
    ];
    
    \node at (axis cs:0.5, 10) [font=\small\sffamily\bfseries] {Safe Harmony};

    \node at (axis cs:0.12, 1) [font=\small\sffamily\bfseries] {Preemption};

    \node at (axis cs:0.05, 4) [font=\small\sffamily\bfseries] {Trust};

    \node at (axis cs:0.6, 2) [font=\small\sffamily\bfseries] {Subversion};

    \end{axis}
\end{tikzpicture}
    \caption{Informed Enemies ($W=1$, $\sigma=0.1$)}
    \label{fig:sc1}
\end{figure}

\begin{figure}[ht]
    \centering
    \begin{minipage}{0.48\textwidth}
    \centering
    \begin{tikzpicture}
    \begin{axis}[
        width=0.7\linewidth,       
        scale only axis,
        xlabel={Relative Capability Lead ($\Delta$)},
        ylabel={Catastrophic Cost ($C$)},
        xmin=0, xmax=1,
        ymin=0, ymax=15, 
        samples=200,
        domain=0:1,
        legend pos=north east,
        axis lines=left,
        clip=true
    ]

    \pgfmathsetmacro{\sigma}{0.2}
    \pgfmathsetmacro{\s}{0.85}
    \pgfmathsetmacro{\e}{1.0}
    \pgfmathsetmacro{\B}{(1 - \s) / \sigma}
    \pgfmathsetmacro{\ksafe}{(\e / (1 - \s)) * (1 - exp(-\B))}
    \pgfmathsetmacro{\ksucker}{(\s * \e) / (1 - \s) * (exp(\B) - 1)}

    
    \addplot[name path=LC, red, thick, domain=0:1] 
        {\ksafe * (1 / (1 + exp(-x / \sigma))) - 1};
        
    \addplot[name path=FUB, blue, thick, domain=0:1] 
        {\ksucker * (1 / (1 + exp(x / \sigma))) - 1};

    \addplot[name path=LUB, orange, dashed, domain=0:1] 
        {\ksucker * (1 / (1 + exp(-x / \sigma))) - 1};
        
    \addplot[name path=FC, green!60!black, dotted, thick, domain=0:1] 
        {\ksafe * (1 / (1 + exp(x / \sigma))) - 1};

    \path[name path=top] (axis cs:0,15) -- (axis cs:1,15);
    \path[name path=zero] (axis cs:0,0) -- (axis cs:1,0);

    \addplot[fill=green!20, opacity=0.6] fill between[of=top and LC, soft clip={domain=0:1}];
    \addplot[fill=green!20, opacity=0.6] fill between[of=top and FUB, soft clip={domain=0:1}];

    \addplot[fill=red!20, opacity=0.6] fill between[of=zero and LC, soft clip={domain=0:1}];

    \addplot[yellow!20] fill between [
        of=LC and FUB,
        split,
        every segment no 0/.style={fill=yellow!20},
        every segment no 1/.style={fill=none}
    ];
    \addplot [orange!20] fill between [
        of=LC and FUB,
        split,
        every segment no 0/.style={fill=none},
        every segment no 1/.style={fill=orange!20}
    ];

    \end{axis}
\end{tikzpicture}
    \caption{Moderately Informed Enemies ($W=1$, $\sigma=0.2$)}
    \label{fig:sc2}
\end{minipage}
\hfill
\begin{minipage}{0.48\textwidth}
    \centering
    \begin{tikzpicture}
    \begin{axis}[
        width=0.7\linewidth,       
        scale only axis,
        xlabel={Relative Capability Lead ($\Delta$)},
        ylabel={Catastrophic Cost ($C$)},
        xmin=0, xmax=1,
        ymin=0, ymax=15, 
        samples=200,
        domain=0:1,
        legend pos=north east,
        axis lines=left,
        clip=true
    ]

    \pgfmathsetmacro{\sigma}{0.1}
    \pgfmathsetmacro{\s}{0.85}
    \pgfmathsetmacro{\e}{0.7}
    \pgfmathsetmacro{\B}{(1 - \s) / \sigma}
    \pgfmathsetmacro{\ksafe}{(\e / (1 - \s)) * (1 - exp(-\B))}
    \pgfmathsetmacro{\ksucker}{(\s * \e / (1 - \s)) * (exp(\B) - 1)}

    
    \addplot[name path=LC, red, thick, domain=0:1] 
        {\ksafe * (1 / (1 + exp(-x / \sigma))) - 1};
        
    \addplot[name path=FUB, blue, thick, domain=0:1] 
        {\ksucker * (1 / (1 + exp(x / \sigma))) - 1};

    \addplot[name path=LUB, orange, dashed, domain=0:1] 
        {\ksucker * (1 / (1 + exp(-x / \sigma))) - 1};
        
    \addplot[name path=FC, green!60!black, dotted, thick, domain=0:1] 
        {\ksafe * (1 / (1 + exp(x / \sigma))) - 1};

    \path[name path=top] (axis cs:0,15) -- (axis cs:1,15);
    \path[name path=zero] (axis cs:0,0) -- (axis cs:1,0);

    \addplot[fill=green!20, opacity=0.6] fill between[of=top and LC, soft clip={domain=0:1}];
    \addplot[fill=green!20, opacity=0.6] fill between[of=top and FUB, soft clip={domain=0:1}];

    \addplot[fill=red!20, opacity=0.6] fill between[of=zero and LC, soft clip={domain=0:1}];

    \addplot[yellow!20] fill between [
        of=LC and FUB,
        split,
        every segment no 0/.style={fill=yellow!20},
        every segment no 1/.style={fill=none}
    ];
    \addplot [orange!20] fill between [
        of=LC and FUB,
        split,
        every segment no 0/.style={fill=none},
        every segment no 1/.style={fill=orange!20}
    ];

    \end{axis}
\end{tikzpicture}
    \caption{Informed Moderate Enemies ($W=0.7$, $\sigma=0.1$)}
    \label{fig:sc3}
\end{minipage}
\end{figure}

Figures \ref{fig:sc1}–\ref{fig:sc3} map the four strategic worlds as an interaction between the capability gap and catastrophic risk, delineated by four critical threshold functions. The Cooperation Thresholds (Green/Red, for Frontrunner and Laggard respectively) determine the cost of loss of control at which the Frontrunner or Laggard switches from racing to pausing, \textit{provided the rival is also pausing}. In contrast, the Unilateral Break Thresholds (Blue/Orange, for Frontrunner and Laggard respectively) identify the point at which a player chooses to pause \textit{if the rival continues to race}.

Strategically, any coordinate located above a player's threshold indicates that the catastrophic risk is high enough to make a pause their rational best response. The Safe Harmony (Green) world exists where both players sit above their respective thresholds and prioritize safety. In the Trust Game (Yellow), the players’ incentives overlap such that mutual pausing is preferred but requires coordination to avoid the sucker's payoff. The world of Subversion (Orange) represents an asymmetric state where the Frontrunner finds it rational to pause while the Laggard still perceives a higher utility in racing to close the gap. Finally, Preemption (Red) occurs when both players fall below their thresholds, meaning the fear of being overtaken or suckered outweighs the fear of catastrophe, forcing a mutual and potentially ruinous race.

The sizes of the possible worlds are moderated by the size of the winner's advantage and the level of technological uncertainty. First, in our model, in line with \cite{armstrongRacingPrecipiceModel2016, emery-xuUncertaintyInformationRisk2024}, we find that a decrease in the size of the winner's advantage makes space for policy to introduce a mutual pause on ASI to become stable (compare Figure \ref{fig:sc1} and \ref{fig:sc2}). If the size of the winner's advantage decreases, then the threshold for the cost of loss of control goes down. Secondly, the model shows that certainty about the possibility of developing ASI encourages racing. In Figure \ref{fig:sc1} (low $\sigma$), the boundaries are sharp. When states are confident it is possible to develop ASI and they value the cost of catastrophe sufficiently low, the Laggard becomes desperate, and the Frontrunner becomes self-confident. Moderate uncertainty ($\sigma=0.2$), as depicted in Figure \ref{fig:sc2}, actually smears the thresholds, expanding the zone where a pause is rational because neither side is certain it is even possible to win the sprint if they choose to race.

\section{The Increasing Likelihood of a Stable Moratorium}

In the previous section, we demonstrated that a moratorium-strategy optimizes a state's self-interest when the cost of loss of control (C) rises sufficiently high, relative to the other parameters. However, decisions are not made based on reality, but the players' perception of it. Hence the question: What about the \textit{perception} of the cost of loss of control? In what direction is it evolving?

A series of expert statements, governmental actions, and public opinion surveys all point in the same direction: the perceived cost of uncontrolled AI is rising, and with it, the strategic space for cooperation. In March 2023, the Future of Life Institute published an open letter calling for a six-month pause on the training of AI systems more powerful than GPT-4. The letter attracted over $30,000$ signatories, including Yoshua Bengio, Stuart Russell, Elon Musk, and Steve Wozniak \cite{PauseGiantAI}. Notably, the letter explicitly warned of race-to-the-bottom dynamics in which labs are incentivized to cut safety corners to deploy products more quickly, which describes the Preemption world in our model. Just two months later, the Center for AI Safety released a one-sentence statement declaring that mitigating the risk of extinction from AI should be treated as a global priority alongside pandemics and nuclear war. This statement was signed by hundreds of leading researchers, including Turing Award laureates Geoffrey Hinton and Yoshua Bengio, as well as the CEOs of OpenAI, Anthropic, and DeepMind \cite{StatementAIRisk}.

Public opinion data suggest that these concerns extend well beyond the research community. A YouGov poll from 2023 found that $43\%$ of respondents were at least somewhat concerned about AI causing humanity’s extinction \cite{AIDoomsdayWorries}. Furthermore, about $60\%$ supported the FLI’s call for a six month pause on AI development\cite{instituteUSPublicWants2025}. A Reuters/Ipsos poll corroborated these findings, as $61\%$ of Americans believed that AI poses risks to humanity \cite{tongAIThreatensHumanitys2023}.

Increased concern among researchers and the public may shift the discourse, but it does not necessarily translate into a corresponding shift in policy-making. However, in the case of ASI, the expert warnings also triggered rapid political action. In October 2023, the Biden administration signed the first major executive order specifically targeting AI safety, requiring developers to share safety test results with the federal government. The following month, the UK government hosted the first-ever global AI safety summit at Bletchley Park. Twenty-eight countries signed the Bletchley Declaration, which raised concerns about catastrophic risks from frontier AI systems \cite{BletchleyDeclarationCountries}. Furthermore, the summit catalyzed the creation of dedicated safety institutions. The UK established its AI Safety Institute in November 2023, with the US following through a parallel body housed within NIST \cite{pillayWhyUSLaunched2024}. By the Seoul AI Summit in May 2024, eleven countries and the European Union had agreed to form an international network of AI Safety Institutes, and South Korea announced its own AI Safety Research Center \cite{WorldLeadersAgree}.

This progression, moving from expert alarm in 2023 to institutional build-out by 2025, signals that the perceived cost of loss of control is a concrete driver of state policy. What was once a fringe concern within the AI safety community has, in under three years, become a mainstream political and institutional priority. This trajectory suggests that, given current perceptions and keeping the other variables stable, the conditions under which a moratorium could plausibly be in a state’s self-interest may be closer than is commonly assumed.

\section{Limitations and Extensions}

In game theory one must make trade-offs between simplicity and realism. We need to carefully select which strategic factors to consider, while also being able to understand their impact. Inevitably, then, one leaves important information out. In future work, the following assumptions could be altered, which may affect the game:

\begin{itemize}
    \item \textit{Dynamic game}. Our model assumes the competition towards ASI is played in one shot. However, it is more realistic to assume the game is played dynamically, such that the game consists of a sequence of moves until a player develops ASI, where the development of ASI becomes more likely through time. This temporal dimension might influence the rationality of the moratorium strategy. For example, we can expect pausing to be rational until the marginal gains from additional safety become lower than the marginal gains from preempting the opponent \cite{lippmanPreemptionRDRaces1988}.
    \item \textit{n-player game}. Our model assumes the game to consist of two players. Although in current circumstances this is a legitimate simplification, more states might build significant capacity to develop ASI in the future. It is thus interesting to extend our two-player game to an $n$-player game in future work. From Armstrong et al.'s paper \cite{armstrongRacingPrecipiceModel2016}, we can expect the presence of more players to have a negative effect on the size of the worlds where a moratorium strategy is rational. 
    \item \textit{n-ary choice}. In our model we assumed a binary policy choice to be made. Although policy choice is coarse grained, hence a discrete set of strategies is more realistic than a continuous one, it would be more realistic to add a third strategy to each player's strategy set. This extra option could be an increase of a state's focus on safety, while continuing the development of ASI. We might expect a `Goldilocks-zone' to emerge where, under very specific circumstances, this strategy is rational.
    \item \textit{Relative capability uncertainty}. It would be interesting to introduce a variable measuring the uncertainty of each player about the relative capability gap, that is, about how much in front or behind they are. We conjecture that, in line with Armstrong et al.\ \cite{armstrongRacingPrecipiceModel2016}, more relative capability uncertainty increases the size of the worlds where a moratorium is stable.
    \item \textit{Beliefs}. In our model we assume the cost of unaligned ASI as an exogenous objective variable. However, in future work, it might be interesting to introduce each player's, potentially asymmetric belief of the cost. We conjecture that different beliefs about the cost of unaligned ASI and the accessibility of information about the other's beliefs will have effects on the rationality of the moratorium strategy.
    \item \textit{Nested games}. We treat nations themselves as rational actors. In reality, policy decisions arise through the actions of a large number of agents within each nation. And while there are nation-level games being played, there are also countless nested games among the actors within the nations. For instance, the US has a two-party system with a four-year presidential election cycle. The ruling party is thus playing a strategic game to retain executive power. This lends importance to short-term policy decisions, which may contradict nation-level long-term rational decisions. 
    \item \textit{Multiple Development}. In our model, we assume the race stops when one of the players develops ASI. This grounds the assumption that the safety-level of the winning player determines the probability of loss of control. However, in reality a player who lost the game might still try to develop their own ASI, which would have an impact both on the strategic choice of each player and the probability of loss of control. In future work, our model can be extended to account for a race that can continue after one player develops ASI.
\end{itemize}

\noindent These and other complications can make for a deeper understanding of the dynamics of ASI-related rational choice, though their addition should be weighed against possible increases in model opacity and decreases in broad applicability, both of which can be side effects of increased complexity.

\section{Conclusion}

This paper has demonstrated that a moratorium on the development of ASI is not necessarily contrary to a state's self-interest, contrary to what critics often claim. Through game-theoretic modeling, we identified four possible strategic worlds: Safe Harmony, Trust, Subversion, and Preemption. The crucial variable is the estimated cost of losing control over an unaligned ASI. When states assess that the potential catastrophic consequences, ranging from permanent loss of human autonomy to extinction, exceed the geopolitical benefits of technological supremacy, relative to technological uncertainty and the safety level, strategic space emerges for a moratorium. Our analysis shows that a smaller size of the winner's advantage between states and a degree of uncertainty about the possibility of developing ASI increases the self-interested basis for cooperation. Finally, based on empirical indicators since 2023, we conclude that although the current geopolitical reality remains dominated by competitive dynamics, the conditions under which a moratorium becomes in a state's self-interest may be closer than commonly assumed.

\section*{Author Contributions}

Edward Roussel: Conceptualization, Methodology, Formal Analysis, Writing - Original Draft. Lode Lauwaert: Conceptualization, Methodology, Writing - Original Draft. Torben Swoboda: Conceptualization, Writing - Original Draft. Grant Ramsey: Conceptualization, Writing - Original Draft. Risto Uuk: Conceptualization, Writing - Reviewing and Editing, Leonard Dung: Conceptualization, Writing - Reviewing and Editing. Anthony Aguirre: Conceptualization, Writing - Reviewing and Editing.

\section*{Acknowledgements}

We would like to thank Prof.\ Dr.\ Sylvia Wenmackers and Prof.\ Dr.\ Luc Lauwers
for their feedback on the mathematical model grounding this paper.
We also thank the members of the Chair Ethics and AI at KU Leuven
and the Future of Life Institute for their feedback and discussion on drafts of this paper. This research was funded by the Research Foundation Flanders (FWO), Grant No.\ 1101426N.

\bibliographystyle{plain}
\bibliography{references.bib}

@misc{grossiWhatAreOdds2025,
	title = {What are the odds? {Risk} and uncertainty about {AI} existential risk},
	shorttitle = {What are the odds?},
	urldate = {2026-05-05},
	publisher = {arXiv},
	author = {Grossi, Marco},
	month = oct,
	year = {2025},
}

@book{goldsteinInternationalRelationsGlobal2020,
	address = {Harlow, UK},
	edition = {twelfth},
	title = {International {Relations}},
	language = {en},
	publisher = {Pearson},
	author = {Goldstein, Joshua S. and Pevehouse, Jon C.},
	year = {2020},
}

@book{bostromSuperintelligencePathsDangers2014,
	address = {Oxford, New York},
	title = {Superintelligence: {Paths}, {Dangers}, {Strategies}},
	isbn = {978-0-19-967811-2},
	shorttitle = {Superintelligence},
	publisher = {Oxford University Press},
	author = {Bostrom, Nick},
	month = jul,
	year = {2014},
}

@misc{uukTaxonomySystemicRisks2024,
	title = {A {Taxonomy} of {Systemic} {Risks} from {General}-{Purpose} {AI}},
	url = {http://arxiv.org/abs/2412.07780},
	doi = {10.48550/arXiv.2412.07780},
	urldate = {2026-03-30},
	publisher = {arXiv},
	author = {Uuk, Risto and Gutierrez, Carlos Ignacio and Guppy, Daniel and Lauwaert, Lode and Kasirzadeh, Atoosa and Velasco, Lucia and Slattery, Peter and Prunkl, Carina},
	month = nov,
	year = {2024},
	note = {arXiv:2412.07780 [cs]},
}

@book{aschenbrennerSituationalAwareness2024,
	title = {Situational {Awareness}},
	author = {Aschenbrenner, Leopold},
	year = {2024},
}

@book{russellHumanCompatibleAI2019,
	title = {Human {Compatible}: {AI} and the {Problem} of {Control}},
	isbn = {978-0-241-33524-6},
	shorttitle = {Human {Compatible}},
	language = {en},
	publisher = {Penguin UK},
	author = {Russell, Stuart},
	year = {2019},
}

@article{sundaramExistentialRiskGlobal2025,
	title = {Existential {Risk} and {Global} {Catastrophic} {Risk}: {A} {Review}},
	shorttitle = {Existential {Risk} and {Global} {Catastrophic} {Risk}},
	url = {https://www.repository.cam.ac.uk/handle/1810/384212},
	doi = {10.17863/CAM.118285},
	language = {eng},
	urldate = {2026-03-19},
	author = {Sundaram, Lalitha and Mani, Lara},
	month = may,
	year = {2025},
}

@article{harackVerificationInternationalAI2025,
	address = {Oxford},
	title = {Verification for {International} {AI} {Governance}},
	journal = {AI Governance Initiative, Oxford Martin School},
	author = {Harack, Ben and Trager, Robert F. and Reuel, Anka and Manheim, David and Brundage, Miles and Aarne, Onni and Scher, Aaron and Pan, Yanliang and Xiao, Jenny and Loke, Kristy},
	year = {2025},
}

@misc{instituteUSPublicWants2025,
	title = {The {U}.{S}. {Public} {Wants} {Regulation} (or {Prohibition}) of {Expert}‑{Level} and {Superhuman} {AI}},
	url = {https://futureoflife.org/recent-news/americans-want-regulation-or-prohibition-of-superhuman-ai/},
	language = {en-US},
	urldate = {2026-03-19},
	journal = {Future of Life Institute},
	author = {{Future of Life Institute}},
	month = oct,
	year = {2025},
}

@techreport{guestBridgingArtificialIntelligence2024,
	title = {Bridging the {Artificial} {Intelligence} {Governance} {Gap}: {The} {United} {States}' and {China}'s {Divergent} {Approaches} to {Governing} {General}-{Purpose} {Artificial} {Intelligence}},
	shorttitle = {Bridging the {Artificial} {Intelligence} {Governance} {Gap}},
	url = {https://www.rand.org/pubs/perspectives/PEA3703-1.html},
	language = {en},
	urldate = {2026-03-19},
	institution = {RAND Corporation},
	author = {Guest, Oliver and Wei, Kevin},
	year = {2024},
}

@article{reynoldsBenchmarkingPathInternational2025,
	title = {Benchmarking as a {Path} to {International} {AI} {Governance}},
	url = {https://www.csis.org/analysis/benchmarking-path-international-ai-governance},
	language = {en},
	urldate = {2026-03-19},
	publisher = {Center for Strategic and International Studies},
	author = {Reynolds, Ian},
	year = {2025},
}

@techreport{ActionPlanIncrease,
	title = {Action {Plan} to {Increase} the {Safety} and {Security} of {Advanced} {AI}},
	url = {https://www.gladstone.ai/action-plan#action-plan-request},
	urldate = {2026-03-19},
	institution = {Gladstone AI},
	author = {Harris, Jeremy and Harris, Edouard and Beall, Mark},
	year = {2023},
	pages = {128},
}

@incollection{mullerFutureProgressArtificial2016,
	address = {Cham},
	title = {Future {Progress} in {Artificial} {Intelligence}: {A} {Survey} of {Expert} {Opinion}},
	isbn = {978-3-319-26485-1},
	shorttitle = {Future {Progress} in {Artificial} {Intelligence}},
	url = {https://doi.org/10.1007/978-3-319-26485-1_33},
	doi = {10.1007/978-3-319-26485-1_33},
	language = {en},
	urldate = {2026-03-19},
	booktitle = {Fundamental {Issues} of {Artificial} {Intelligence}},
	publisher = {Springer International Publishing},
	author = {Müller, Vincent C. and Bostrom, Nick},
	editor = {Müller, Vincent C.},
	year = {2016},
	pages = {555--572},
}

@article{graceThousandsAIAuthors2025,
	title = {Thousands of {AI} {Authors} on the {Future} of {AI}},
	volume = {84},
	copyright = {Copyright (c) 2025 Journal of Artificial Intelligence Research},
	issn = {1076-9757},
	url = {https://www.jair.org/index.php/jair/article/view/19087},
	doi = {10.1613/jair.1.19087},
	language = {en},
	urldate = {2026-03-19},
	journal = {Journal of Artificial Intelligence Research},
	author = {Grace, Katja and Sandkühler, Julia Fabienne and Stewart, Harlan and Weinstein-Raun, Benjamin and Thomas, Stephen and Stein-Perlman, Zach and Salvatier, John and Brauner, Jan and Korzekwa, Richard C.},
	month = oct,
	year = {2025},
}

@misc{pethokoukisAIBanBackers2025,
	title = {{AI} {Ban} {Backers} {Risk} {Freezing} {Progress}},
	url = {https://www.aei.org/economics/ai-ban-backers-risk-freezing-progress/},
	language = {en-US},
	urldate = {2026-03-18},
	journal = {American Enterprise Institute},
	author = {Pethokoukis, James},
	month = oct,
	year = {2025},
}

@misc{BletchleyDeclarationCountries,
	title = {The {Bletchley} {Declaration} by {Countries} {Attending} the {AI} {Safety} {Summit}, 1-2 {November} 2023},
	url = {https://www.gov.uk/government/publications/ai-safety-summit-2023-the-bletchley-declaration/the-bletchley-declaration-by-countries-attending-the-ai-safety-summit-1-2-november-2023},
	language = {en},
	urldate = {2026-03-09},
	author = {{UK Government}},
	month = feb,
	year = {2025},
}

@misc{StatementSuperintelligence2025,
	title = {Statement on {Superintelligence}},
	url = {https://superintelligence-statement.org},
	language = {en},
	urldate = {2025-11-26},
	author = {{Future of Life Institute}},
	year = {2025},
}

@misc{StatementAIRisk,
	title = {Statement on {AI} {Risk}},
	url = {https://aistatement.com},
	language = {en},
	urldate = {2026-03-09},
	author = {{Center for AI Safety}},
	year = {2023},
}

@misc{PauseGiantAI,
	title = {Pause {Giant} {AI} {Experiments}: {An} {Open} {Letter}},
	shorttitle = {Pause {Giant} {AI} {Experiments}},
	url = {https://futureoflife.org/open-letter/pause-giant-ai-experiments/},
	language = {en-US},
	urldate = {2026-03-09},
	author = {{Future of Life Institute}},
	year = {2023},
}

@article{tongAIThreatensHumanitys2023,
	chapter = {Technology},
	title = {{AI} {Threatens} {Humanity}’s {Future}, 61\% of {Americans} say: {Reuters}/{Ipsos} poll},
	shorttitle = {{AI} threatens humanity’s future, 61\% of {Americans} say},
	url = {https://www.reuters.com/technology/ai-threatens-humanitys-future-61-americans-say-reutersipsos-2023-05-17/},
	language = {en},
	urldate = {2026-03-09},
	journal = {Reuters},
	author = {Tong, Anna},
	month = may,
	year = {2023},
}

@misc{pillayWhyUSLaunched2024,
	title = {Why the {U}.{S}. {Launched} an {International} {Network} of {AI} {Safety} {Institutes}},
	url = {https://time.com/7178133/international-network-ai-safety-institutes-convening-gina-raimondo-national-security/},
	language = {en},
	urldate = {2026-03-09},
	journal = {TIME},
	author = {Pillay, Tharin},
	month = nov,
	year = {2024},
}

@misc{AIDoomsdayWorries,
	title = {{AI} {Doomsday} {Worries} many {Americans}. {So} does {Apocalypse} from {Climate} {Change}, {Nukes}, {War}, and {More}},
	url = {https://yougov.com/en-us/articles/45565-ai-nuclear-weapons-world-war-humanity-poll},
	language = {en-us},
	urldate = {2026-03-09},
	journal = {YouGov},
	author = {Orth, Taylor and Bialik, Carl},
	month = apr,
	year = {2023},
}

@misc{WorldLeadersAgree,
	title = {World {Leaders} {Agree} to {Launch} {Network} of {AI} {Safety} {Institutes}},
	url = {https://www.euronews.com/next/2024/05/22/ai-seoul-summit-world-leaders-agree-to-launch-network-of-safety-institutes},
	language = {en},
	urldate = {2026-03-09},
	journal = {Euronews},
	author = {Desmarais, Anna},
	month = may,
	year = {2024},
	note = {Section: next\_tech-news},
}

@article{lippmanPreemptionRDRaces1988,
	title = {Preemption in {R}\&{D} races},
	volume = {32},
	issn = {0014-2921},
	number = {8},
	urldate = {2026-03-04},
	journal = {European Economic Review},
	author = {Lippman, Steven A. and McCardle, Kevin F.},
	month = oct,
	year = {1988},
	pages = {1661--1669},
}

@article{jiaContestFunctionsTheoretical2013,
	series = {Tournaments, {Contests} and {Relative} {Performance} {Evaluation}},
	title = {Contest functions: {Theoretical} foundations and issues in estimation},
	volume = {31},
	issn = {0167-7187},
	shorttitle = {Contest functions},
	url = {https://www.sciencedirect.com/science/article/pii/S0167718712000811},
	doi = {10.1016/j.ijindorg.2012.06.007},
	number = {3},
	urldate = {2026-02-24},
	journal = {International Journal of Industrial Organization},
	author = {Jia, Hao and Skaperdas, Stergios and Vaidya, Samarth},
	month = may,
	year = {2013},
	pages = {211--222},
}

@book{ordPrecipiceExistentialRisk2021,
	address = {London},
	title = {The {Precipice}: {Existential} {Risk} and the {Future} of {Humanity}},
	isbn = {978-1-5266-0023-3},
	shorttitle = {The {Precipice}},
	language = {eng},
	publisher = {Bloomsbury Academic},
	author = {Ord, Toby},
	year = {2021},
}

@article{snyderPrisonersDilemmaChicken1971,
	title = {“{Prisoner}'s {Dilemma}” and “{Chicken}” {Models} in {International} {Politics}},
	volume = {15},
	issn = {0020-8833},
	url = {https://doi.org/10.2307/3013593},
	doi = {10.2307/3013593},
	number = {1},
	urldate = {2026-02-03},
	journal = {International Studies Quarterly},
	author = {Snyder, Glenn H.},
	month = mar,
	year = {1971},
	pages = {66--103},
}

@article{jervisCooperationSecurityDilemma1978,
	title = {Cooperation {Under} the {Security} {Dilemma}},
	volume = {30},
	issn = {0043-8871},
	url = {https://www.jstor.org/stable/2009958},
	doi = {10.2307/2009958},
	number = {2},
	urldate = {2026-02-03},
	journal = {World Politics},
	publisher = {[Trustees of Princeton University, The Johns Hopkins University Press]},
	author = {Jervis, Robert},
	year = {1978},
	pages = {167--214},
}

@book{schellingStrategyConflict1963,
	address = {Cambridge (Mass.)},
	edition = {2nd pr.},
	title = {The strategy of conflict.},
	language = {eng},
	publisher = {Harvard university},
	author = {Schelling, Thomas C.},
	year = {1963},
}

@article{hirshleiferConflictRentseekingSuccess1989,
	title = {Conflict and rent-seeking success functions: {Ratio} vs. difference models of relative success},
	volume = {63},
	issn = {1573-7101},
	shorttitle = {Conflict and rent-seeking success functions},
	url = {https://doi.org/10.1007/BF00153394},
	doi = {10.1007/BF00153394},
	language = {en},
	number = {2},
	urldate = {2026-01-26},
	journal = {Public Choice},
	author = {Hirshleifer, Jack},
	month = nov,
	year = {1989},
	pages = {101--112},
}

@book{kyddTrustMistrustInternational2007,
	address = {Princeton, NJ Oxford},
	title = {Trust and {Mistrust} in {International} {Relations}},
	isbn = {978-0-691-13388-1},
	language = {eng},
	publisher = {Princeton University Press},
	author = {Kydd, Andrew H.},
	year = {2007},
}

@misc{dungRacingAGICooperation2025,
	title = {Against racing to {AGI}: {Cooperation}, deterrence, and catastrophic risks},
	shorttitle = {Against racing to {AGI}},
	url = {http://arxiv.org/abs/2507.21839},
	doi = {10.48550/arXiv.2507.21839},
	urldate = {2026-01-19},
	publisher = {arXiv},
	author = {Dung, Leonard and Hellrigel-Holderbaum, Max},
	month = jul,
	year = {2025},
	note = {arXiv:2507.21839 [cs]},
}

@article{emery-xuUncertaintyInformationRisk2024,
	title = {Uncertainty, {Information}, and {Risk} in {International} {Technology} {Races}},
	volume = {68},
	issn = {0022-0027},
	url = {https://doi.org/10.1177/00220027231214996},
	doi = {10.1177/00220027231214996},
	language = {EN},
	number = {10},
	urldate = {2026-01-19},
	journal = {Journal of Conflict Resolution},
	publisher = {SAGE Publications Inc},
	author = {Emery-Xu, Nicholas and Park, Andrew and Trager, Robert},
	month = nov,
	year = {2024},
	pages = {2019--2047},
}

@misc{youngWhosDrivingGame2025,
	title = {Who's {Driving}? {Game} {Theoretic} {Path} {Risk} of {AGI} {Development}},
	shorttitle = {Who's {Driving}?},
	url = {http://arxiv.org/abs/2501.15280},
	doi = {10.48550/arXiv.2501.15280},
	urldate = {2026-01-07},
	publisher = {arXiv},
	author = {Young, Robin},
	month = jan,
	year = {2025},
	note = {arXiv:2501.15280 [cs]},
}

@misc{hendrycksSuperintelligenceStrategyExpert2025,
	title = {Superintelligence {Strategy}: {Expert} {Version}},
	shorttitle = {Superintelligence {Strategy}},
	url = {http://arxiv.org/abs/2503.05628},
	doi = {10.48550/arXiv.2503.05628},
	urldate = {2026-01-07},
	publisher = {arXiv},
	author = {Hendrycks, Dan and Schmidt, Eric and Wang, Alexandr},
	month = apr,
	year = {2025},
	note = {arXiv:2503.05628 [cs]},
}

@techreport{abrahamPrisonersDilemmaRace2026,
	address = {Santa Monica, CA},
	title = {A {Prisoner}’s {Dilemma} in the {Race} to {Artificial} {General} {Intelligence}},
	url = {https://www.rand.org/pubs/research_reports/RRA4245-1.html},
	language = {en},
	urldate = {2026-01-07},
	institution = {RAND Corporation},
	author = {Abraham, Lisa and Kavner, Joshua and Moon, Alvin},
	year = {2026},
}

@misc{katzkeManhattanTrapWhy2025,
	title = {The {Manhattan} {Trap}: {Why} a {Race} to {Artificial} {Superintelligence} is {Self}-{Defeating}},
	shorttitle = {The {Manhattan} {Trap}},
	url = {https://www.ssrn.com/abstract=5067833},
	doi = {10.2139/ssrn.5067833},
	urldate = {2026-01-07},
	publisher = {SSRN},
	author = {Katzke, Corin and Futerman, Gideon},
	year = {2025},
}

@article{armstrongRacingPrecipiceModel2016,
	title = {Racing to the {Precipice}: a {Model} of {Artificial} {Intelligence} {Development}},
	volume = {31},
	issn = {1435-5655},
	shorttitle = {Racing to the precipice},
	url = {https://doi.org/10.1007/s00146-015-0590-y},
	doi = {10.1007/s00146-015-0590-y},
	language = {en},
	number = {2},
	urldate = {2026-01-06},
	journal = {AI \& SOCIETY},
	author = {Armstrong, Stuart and Bostrom, Nick and Shulman, Carl},
	month = may,
	year = {2016},
	pages = {201--206},
}

@book{petersGameTheoryMultiLeveled2015,
	address = {Berlin, Heidelberg},
	title = {Game {Theory}: {A} {Multi}-{Leveled} {Approach}},
	volume = {Second Edition},
	language = {en},
	publisher = {Springer Berlin Heidelberg},
	author = {Peters, Hans},
	year = {2015},
}

\section*{Appendix}

\subsection*{The Model}\label{sec:tm}

We formalize the strategic model described in Section \ref{sec:spf} above as a two player game, with players $i \in \{1, 2\}$. Let $\Delta$ denote the Frontrunner's relative capability advantage. Let $s_i \in [0, 1]$ be the safety policy taken by player $i$, where $s_i=1$ represents a Moratorium (maximum safety) and $s_i \in [0,1)$ represents a Race. In this paper we assume $s=0.85$, if a player races, but in the model we allow different possibilities.

We model the probability of victory as a continuous logistic contest function \cite{hirshleiferConflictRentseekingSuccess1989,jiaContestFunctionsTheoretical2013}, moderated by the level of technological uncertainty $\sigma > 0$. The probability of the Frontrunner winning when both players pause is given by:

\begin{equation*}
\label{eq:p_win}
    P_{win}(\Delta) = P = \frac{1}{1 + e^{-\frac{\Delta}{\sigma}}}.
\end{equation*}

Conversely, the probability of the Frontrunner losing (and the Laggard winning) is:

\begin{equation*} \label{eq:p_lose}
    P_{lose}(\Delta) = 1 - P= \frac{1}{1 + e^{\frac{\Delta}{\sigma}}}.
\end{equation*}

As $\sigma \to 0$, the outcome becomes deterministic, meaning that the smallest advantage in capability results in a sure win, as $\sigma$ increases, there is more uncertainty about the possibility of ASI, resulting in a lower probability of winning.

A player who chooses to Race accepts a safety penalty $s \in [0,1)$ to gain a speed advantage relative to a Pausing rival. We define the \textit{race Boost Constant} ($B$) to capture the normalized strategic value of this acceleration:
\begin{equation*}
    B = \frac{1-s}{\sigma}.
\end{equation*}
This constant allows us to express the complex logistic shifts as simple linear multipliers of the base probabilities.

We define the probability of winning when racing against a pausing opponent ($P^R$) and the probability of winning when pausing against a racing opponent ($P^S$):
\begin{equation*}
    P^R = \frac{P}{P + (1-P)e^{-B}}, \quad P^S = \frac{P}{P + (1-P)e^{B}}.
\end{equation*}

A player gets a payoff $1$ if they win the race safely and $1-W$ if they lose safely, where $W\in[0,1]$ is the size of the winner's advantage in the game. We define the cost of uncontrolled ASI as $C$. Since the utility for winning is normalized to $1$, $C$ is interpreted relative to benefit. Contrary to Armstrong et al.\ \cite{armstrongRacingPrecipiceModel2016}, we allow $C > 1$, reflecting that the disutility of loss of control exceeds the utility of winning. The probability of a safe outcome is endogenous to the strategy; a winning player with strategy $s_i$ avoids catastrophe with probability $s_i$.

Finally, the expected utility function for Player $i$ is:
\begin{equation*}
    EU_i(s_i, s_j) = P(i\ \text{Win}) \cdot [s_i - (1-s_i)C] + P(i\ \text{Lose}) \cdot [s_j(1-W) - (1-s_j)C].
\end{equation*}

\subsection*{Derivation of Strategic Thresholds}\label{sec:dst}

To derive the thresholds governing the stable outcomes, we identify the indifference thresholds where the expected utility ($EU$) of a Moratorium ($s_i = 1$) equals the utility of a Race ($s_i = s$). Let $P$ denote the base win probability for the player in question.

A critical identity used in these derivations is the ratio of base probability to shifted probability:
\begin{equation*}
    \frac{P}{P^R} = P + (1-P)e^{-B}, \quad \frac{P}{P^S} = P + (1-P)e^{B}.
\end{equation*}

We can now derive the Frontrunner's thresholds.
Remember that $P = P_{win}(\Delta)$ and $P_{lose}(\Delta) = 1-P$, the win, respectively lose, probability for the Frontrunner if both play the same strategy.

\paragraph{Frontrunner Cooperation (FC):}
The Frontrunner pauses if $EU_L(1, 1) > EU_L(s, 1)$. 
\begin{equation}
    P + (1-P)(1-W) \geq P^R[s - (1-s)C] + (1-P^R)(1-W)
\end{equation}
Rearranging for $C$:
\begin{equation}
    P^R(1-s)C \geq e(P^R - P) - P^R(1-s) \implies C \geq \frac{W}{1-s}\left(1 - \frac{P}{P^R}\right) - 1.
\end{equation}
Substituting the identity $1 - \frac{P}{P^R} = (1-P)(1-e^{-B})$:
\begin{equation}
    \textbf{FC}: C \geq \frac{W}{1-s}(1 - e^{-B})P_{lose}(\Delta) - 1.
\end{equation}

\paragraph{Frontrunner Unilateral Break (FUB):}
The Frontrunner pauses if $EU_L(1, s) \geq EU_L(s, s)$.
\begin{equation}
    P^S + (1-P^S)[s(1-W) - (1-s)C] \geq P[s - (1-s)C] + (1-P)[s(1-W) - (1-s)C]
\end{equation}
Factoring the terms yields:
\begin{equation}
    P^S(1-s)C \geq (P - P^S)(s - 1 - sW) \implies C \geq \left(\frac{P - P^S}{P^S}\right)\left(\frac{s - 1 - sW}{1 - s}\right) - 1
\end{equation}
Using $\frac{P}{P^S} - 1 = (1-P)(e^B - 1)$:
\begin{equation}
    \textbf{FUB}: C \geq \frac{sW}{1-s}(e^B - 1)P_{lose}(\Delta) - 1
\end{equation}

The Laggard's thresholds are derived symmetrically (since $\Delta < 1$). Let $P = P_{lose}(\Delta)$, which is identical to $P_{win}$ for the Frontrunner. By the symmetry of the logistic function, $1-P = P_{win}(\Delta)$.

\paragraph{Laggard Cooperation (LC):}
Following the derivation of FC but substituting the Laggard's base win probability:
\begin{equation}
    \textbf{LC}: C \geq \frac{W}{1-s}(1 - e^{-B})P_{win}(\Delta) - 1
\end{equation}

\paragraph{Laggard Unilateral Break (LUB):}
Following the derivation of FUB but substituting the Laggard's base win probability:
\begin{equation}
    \textbf{LUB}: C \geq \frac{sW}{1-s}(e^B - 1)P_{win}(\Delta) - 1
\end{equation}

\end{document}